\documentclass[intlimits,twoside,a4paper]{article}
\usepackage{graphicx}
\usepackage[cp1251]{inputenc}

\usepackage{cmpj3}

\issue{2017}{20}{3}{33004}
\doinumber{10.5488/CMP.20.33004}

\title[Charge storage in nanotubes: the case of a 2-1 electrolyte]{Charge storage in nanotubes: the case of a 2-1~electrolyte}
\author[W. Schmickler, D. Henderson]{W. Schmickler\refaddr{label1}, D. Henderson\refaddr{label2}}
\addresses{
\addr{label1}Institute of Theoretical Chemistry, Ulm University, Germany
\addr{label2}Department of Chemistry and Biochemistry, Brigham Young University, Provo, UT 84602, USA
}

\date{Received July 25, 2017, in final form August 7, 2017}
\begin{document}
\maketitle

\begin{abstract}
We consider a 2-1 electrolyte in contact with a narrow nanotube, which only allows one-dimensional  storage along the axis. The asymmetry does not allow
an a priori definition of the potential of zero charge; instead, the natural reference is the electrode potential at which both ions have the same electrochemical potential; the value of the latter can serve as a measure of ionophilicity. Near this potential, ionophobic tubes  are filled with a dilute gas, ionophilic tubes are filled with a one-dimensional solid containing about the same number of the divalent ions and the monovalent counterions, a structure that is stabilized by a strong screening of the Coulomb interaction by an induced counter charge  on the walls of the tube.
The filling of the tube by the application of an electrode potential exhibits  a complicated pattern of interactions between the two kinds of ions.

\pacs 02.70.Lq, 82.20.Wt, 82.45.+z, 83.20
\end{abstract}

\section{Introduction}
We met Jean-Pierre Badiali first at a workshop on the electrochemical double layer, which was organized by Roger Parsons
at the CNRS Laboratory Bellevue, Paris, in 1982. To our surprise, we found out that we  were working on similar models for the double layer,
combining the jellium model for the metal with the hard-sphere electrolyte model for the solution \cite{us1,badi1,badi12}. Subsequently, 
we worked in parallel on this topic, exchanging ideas openly, but never writing a joint publication. This only came about much later, when one of us (WS) worked with him on another topic, on electron transfer reactions \cite{jpme}. 

Jean-Pierre worked on a broad spectrum of topics in physics, some of them quite fundamental, and it was always a challenge, an intellectual and personal pleasure  to discuss science with him.  But the double layer is the topic which brought us together, so we thought it fit to dedicate an article on a  modern double layer problem to his memory. 

Nowadays there is much interest in double layers in confined spaces, whose extension is smaller than the Debye length of the solution.
An extreme case is a nanotube which is so thin that only one line of ions can enter. When such a nanotube serves as an electrode and is in contact with an ionic liquid 
or a molten salt, ions can enter and form an electric double layer, whose composition can be controlled by the electrode potential.
This case might be thought to be of academic interest only, but there is a practical side to it: The storage of ions in nanotubes is of great practical importance for supercapacitors and for batteries, 
and narrow tubes with diameters of less than 1~nm have an unusually high capacitance per area \cite{ex1,ex2}.
Therefore, much attention has been focused on the storage of a one-dimensional line of ions inside a narrow tube, e.g., \cite{ising1,ising2,mypore,roch,doug}. 

In a very recent work \cite{pap1} we have investigated a 1-1 electrolyte in contact with a nanotube by grand-canonical Monte Carlo simulations. In that work, the focus was on the interfacial capacitance. Strangely, the case of an asymmetrical electrolyte seems to have been neglected so far.
So, in this work we consider a 2-1 electrolyte in a nanotube, and focus on the consequences which the composition has on the structure
of the embedded chain of ions, and on the stored charge as a function of potential.

\section{Screening of the Coulomb interaction and the formation of a one-dimensional salt}
The total system, i.e., incorporated ions and induced charge on the walls of the tube, is always neutral. An ion embedded in the tube generates an image charge
of equal magnitude and opposite sign, which spreads in a ring-like manner around the ion.\footnote{There is a moot discussion in the community whether to call this an image charge or an induced charge. We follow our previous publications and use both expression as synonymous. In fact, the optical image of a sphere embedded in a reflecting tube is a ring.} When the ion is on the axis of the tube, the electrostatic problem can be solved for a perfectly conducting, classical tube. In particular, along the axis of the tube, the electrostatic potential generated by the ion and its image charge is then given by \cite{jackson,kondrat}:
\begin{equation}
\label{clas}
\Phi(z) =\frac{2}{R}\ \sum_{m=1}^{\infty} \frac{\exp(-k_mz/R)}{k_m|J_1(k_m)|^2} \,,
\end{equation}
where $R$ is the radius of the tube, $z$ is the distance from the center of the ion, $k_m$ denotes the roots of the Bessel function $J_0(k_m)=0$, and $J_1$ is the Bessel function of first order. 

\begin{figure}[!h]
\vspace{-3mm}
\begin{center}
  \includegraphics[height=6cm]{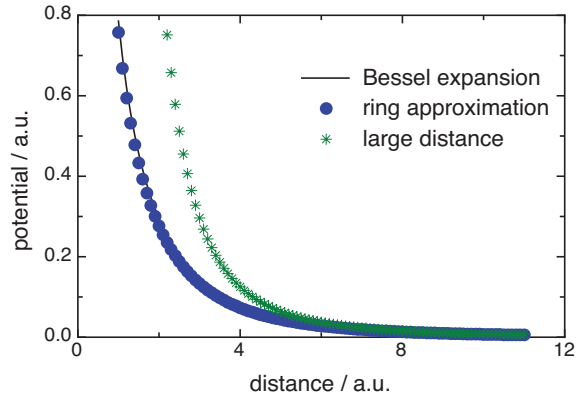} 
      \end{center}
  \caption{(Color online) Potential along the axis of a ring with an effective image radius of 4 a.u. The full line corresponds to exact solution of equation~(\ref{clas}),
  the ring approximation is given by equation~(\ref{dip}), and the large distance approximation refers to the $1/z^3$ asymptote.}   
     \label{potent}
\end{figure}

A real tube, such as a carbon or gold nanotube, screens the charge even better than a perfect classical metal, since the electrons spill over the metal surface. Mohammadzadeh et al. \cite{me1,me2,me3,me4} have shown that in this case equation~(\ref{clas}) can still be used, but the classical radius 
must be replaced by an effective radius, which is typically smaller by about 0.5~\AA\ than the physical radius defined by the positions of the atoms. Figure~\ref{potent} shows the potential along the axis for a typical case.

Formula~(\ref{clas}) is of little practical use in Monte-Carlo or molecular dynamics simulations,
since its evaluation requires too much time, and for large distances one needs a large number of terms. A useful approximation is based
 on a physical consideration: The image charge forms a ring around the ion. At large distances, the width of this ring can be neglected; the resulting potential is:
\begin{equation}
\label{dip}
\Phi(z) \approx \frac{1}{z} - \frac{1}{\sqrt{z^2+R^2}} \approx \frac{R^2}{2z^3}\,.
\end{equation}
The latter approximation holds when $|z|\gg R$; it shows that at large distances the interaction decreases with $1/z^3$, while for an unscreened  Coulomb potential it decreases with  $1/z$.  Figure~\ref{potent} compares the exact solution with the approximate expression of equation~(\ref{dip}) and the asymptotic form. A  tube filled with one kind of ion, all of the same charge, can be stable, since:

\begin{equation}
\label{ }
\sum_{n=1}^\infty \frac{1}{n^3} = \zeta (3) \approx 1.20205,
\end{equation}
where $\zeta(x)$ is the Riemann $\zeta$-function.
The corresponding sum $\sum_{n=1}^\infty 1/n$ for the unscreened Coulomb potential diverges.  One has to keep in mind that the total system, i.e., ions plus nanotube, is always uncharged. Of course, this holds only for nanotubes with metallic conductivity. In this context it is noteworthy that semiconducting carbon nanotubes may become conductive by the insertion of ions \cite{me2}.

The asymptotic behavior allows us to estimate which structure is stable for the case of a 2-1 electrolyte lined up along the axis of the tube.
There are two obvious candidates for a divalent anion, univalent cation:
\begin{center}
 \includegraphics[width=0.9\textwidth]{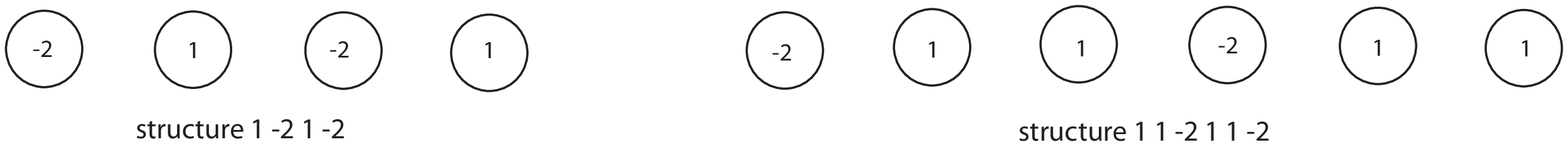} 
      \end{center}
In order to find out which of the two is more stable, we calculate the following lattice sums, setting the lattice constant equal to unity:      
interaction of one anion with all other ions plus interaction of one cation with all other ions minus the interaction of a neighboring anion-cation pair, the latter to avoid double counting. The sums are all related to $\sum 1/n^3$. The corresponding values are: $-4.9118$ for the 1~-2~1~-2 structure, and
$-3.3423$ for the 1~1~-2~1~1~-2 structure. If we restrict ourselves to nearest neighbor interactions only, the values are $-6$ 
for the 1~-2~1~-2 structure, and
$-5$ for the 1~1~-2~1~1~-2 structure. Thus, the former is more stable, even though the unit cell carries an excess charge. This estimate will be confirmed by our Monte Carlo simulations.

\section{Grand canonical Monte Carlo simulations}
\begin{figure}[!t]
\vspace{-5mm}
\begin{center}
  \includegraphics[height=6cm]{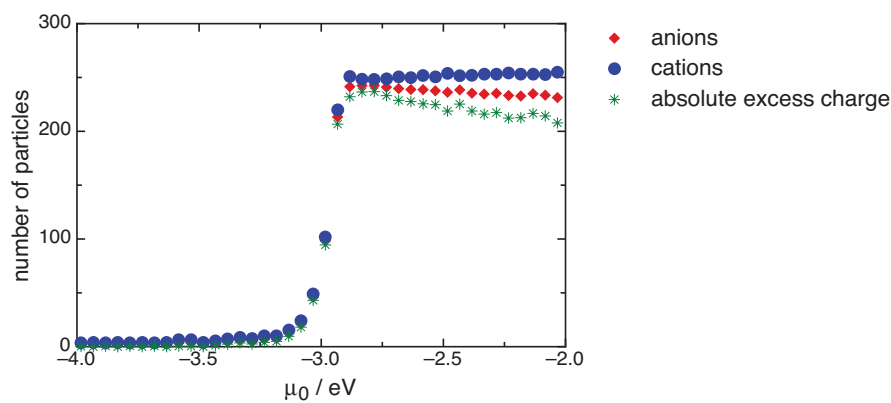} 
      \end{center}
      \vspace{-3mm}
  \caption{(Color online) Number of particles as a function of the ionophilitity.}   
     \label{fillt}
\end{figure}

\begin{figure}[!t]
\vspace{-12mm}
\begin{center}
  \includegraphics[height=3cm]{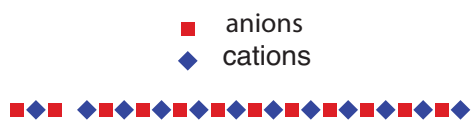} 
      \end{center}
  \caption{(Color online) Snapshot of the particle arrangement for a tube that is almost filled. In this example there is one vacancy for reasons of entropy.}   
     \label{snap}
\end{figure}

For a 1-1 electrolyte, the potential of zero charge is an obvious reference potential. It coincides with the potential at which
both ions have the same electrochemical potential. For an asymmetrical electrolyte there is no obvious way to determine the potential of zero charge a priori, but the potential   at which both ions have the same electrochemical potential $\mu_0$ can still be defined and can serve as the reference. Thus, we have for our (1,-2) electrolyte:
\begin{equation}
\label{mus}
\mu_+ = \mu_0 + e_0 \Delta \phi,\qquad \mu_- = \mu_0 -2 e_0 \Delta \phi.
\end{equation}
An important question is how the the nanotube is filled at the reference potential when the electrochemical potential $\mu_0$ changes. With our sign convention \cite{pap1} increasing $\mu_0$ corresponds to increasing ionophilicity. The results are shown in figure~\ref{fillt}.
For low values of $\mu_0$, the tube is almost empty, the occupation corresponds to a dilute gas. At a critical value near
$\mu_0 \approx -3.2$~eV, the occupation rises rapidly. Just as in the case of a 1-1 electrolyte in this region, the fluctuations are large, and the transition region has a finite width, because there is no phase transition in one-dimensional systems. 
Above $\mu_0 \approx -3.0$~eV the tube is filled with a solid structure, i.e., a one-dimensional salt. As predicted above, the number of 
anions and cations is almost equal, so that the structure corresponds roughly to 1~-2~1~-2. A typical snapshot from a tube that is almost filled is seen in figure~\ref{snap}. This implies that the negative excess charge is nearly equal to the number of anions. With an increasing $\mu_0$, the number of cations becomes slightly larger at the expense of the anions, so that the negative excess charge is somewhat reduced. In accord with the terminology of Lee et al. \cite{ising1} we call the tubes that are almost empty at $\mu_0$ ionophobic, those that are filled as ionophilic; the transition region between the two cases is small. 
\begin{figure}[!t]
\begin{center}
  \includegraphics[height=6cm]{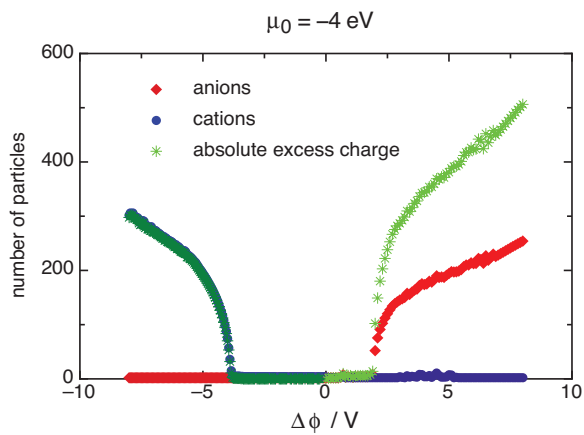} 
      \end{center}
       \vspace{-3mm}
  \caption{(Color online) Filling of the nanotube as a function of potential for an ionophobic tube with $\mu_0 = -4$~eV.}   
     \label{badi4}
\end{figure}
In a typical experiment, the ionophilicity, which depends on the material of the tube and on the electrolyte, is kept constant, while the
electrode potential $\Delta \phi$ is scanned. The corresponding charging curves depend on $\mu_0$. We first consider the case where
the tube is strongly ionophobic, choosing $\mu_0 = -4$~eV in accord with figure~\ref{fillt}. The results are shown in figure~\ref{badi4}. On a positive scan,  the tube starts to fill with anions with a certain delay. At first the number of particles rises rapidly, but then the Coulomb repulsion between the ions becomes noticeable, and the rate of filling slows down. 
On a negative scan, the filling with cations requires a larger absolute value of $\Delta \phi$ since they carry only a single unit of charge. 
Again, at first the tube is filled rapidly; then, the rate decreases, but not as strongly  as in the positive sweep, since the charges on the ions are smaller and hence the repulsion is weaker. All in all, the behavior is similar to that of an ionophobic tube in contact with a 1-1 electrolyte \cite{pap1}.
\begin{figure}[!t]
\vspace{-5mm}
\begin{center}
  \includegraphics[height=6cm]{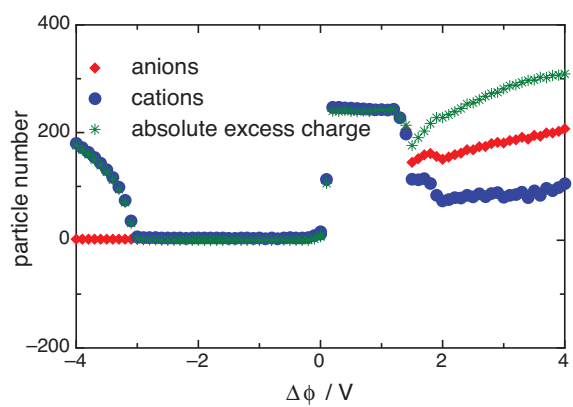} 
      \end{center}
  \caption{(Color online) Filling of the nanotube as a function of potential for an ionophobic tube with $\mu_0 = -3.2$~eV.}   
     \label{g32}
\end{figure}

The filling is most interesting if we start with $\mu_0$ near the transition region between the gas-like and the solid-like region. Figure~\ref{g32} shows the case with $\mu_0 = -3.2$~eV. The negative scan is similar to the previous case with $\mu_0 = -4$~eV:
On the application of a negative potential, the tube becomes empty and stays so, till at about $\Delta \phi = -3$~V the cations start to enter, at first rapidly, and then with a smaller rate. The positive scan is much more interesting: At first anions and cations enter in equal numbers, the tube is filled almost completely, and a 
one-dimensional solid is formed. With an increasing potential, the energetics of the cations becomes less and less favorable. Suddenly, cations start to leave and a few anions follow them.The excess negative charge also drops, but at higher potentials  more anions enter, and the charge rises accordingly. Nevertheless, the anions that enter drag a few cations with them, but the negative excess charge keeps on rising. A complicated interplay between the two kinds of ions!

\begin{figure}[!b]
\vspace{-3mm}
\begin{center}
  \includegraphics[height=6cm]{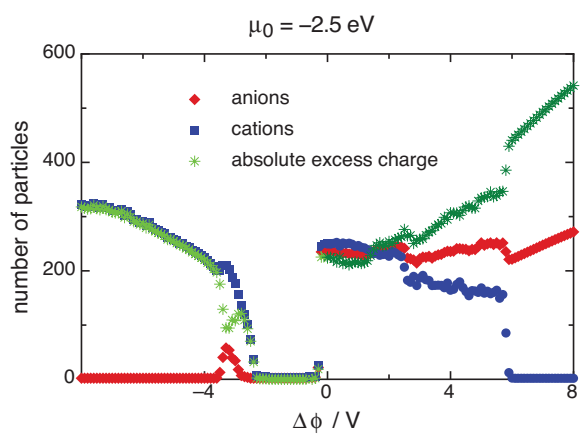} 
      \end{center}
  \caption{(Color online) Filling of the nanotube as a function of potential for an ionophobic tube with $\mu_0 = -2.5$~eV.}   
     \label{g25}
\end{figure}
A similar complicated behavior is observed for an even more ionophilic tube with $\mu_0 = -2.5$~eV (see figure~\ref{g25}). 
For $\Delta \phi = 0$, the tube is almost completely filled, the number of cations being slightly larger than that of the anions --- this is the same situation that we observed in figure~\ref{g32}. With an increasing $\Delta \phi$,  the cations start to leave, at first slowly, till their number suddenly drops to zero. The number of anions at first stays roughly constant, till their number starts to  rise when all the cations have left.

During a negative sweep of $\Delta \phi$, all ions at first leave. Then, the number of cations at first rises rapidly, and then more slowly as the repulsion becomes noticeable. There is a curious small hump in the number of cations at the potential where the curve for the cations changes its slope. This effect had also been observed for the 1-1 electrolyte \cite{pap1}.

Finally, we note that figure~\ref{badi4} illustrates an idea of Kondrat and Kornyshev \cite{spring}: Energy storage is most effective in an ionophobic tube, which can just be completely filled at the limits of the experimentally accessible potential. 

\section{Conclusions}
An asymmetrical 2-1 electrolyte exhibits a number of intriguing features. Firstly, there is no unique potential of zero charge. For a symmetrical  electrolyte, this is the potential where both kinds of ions have the same electrochemical potential. As shown in figure~\ref{fillt}, only for ionophobic tubes with a low $\mu_0$, the tube is uncharged --- in fact it is almost empty. However, in this case it is uncharged over a range of potentials (see figure~\ref{badi4}). By contrast, ionophilic tubes are  filled with an almost equal number of cations and anions  at $\Delta \phi=0$, and thus carry an excess charge. Ionophilic tubes have an extended range of zero charge at negative potentials, where the doubly-charged anions are driven out, and the cations do not yet enter.

Particularly, at positive potentials the ionophilic curves show a complicated filling pattern: At first, both types of ions are present, and then the counter-ions start to leave. This expulsion tends to set in rather abruptly, and in this transition region the total charge may even drop slightly with an increasing potential, giving rise to a negative capacitance.

As we have stated at the beginning, such one-dimensional storage is difficult to realize experimentally, so  our work should be viewed as a contribution to the study of a model system, which has been discussed quite extensively in the literature. 

\section*{Acknowledgements} W.S. thanks the  University of Science and Technology of China, Hefei, and particularly his host, Prof.~YanXia Chen, for a visiting professorship. In addition he gratefully acknowledges financial support by the Deutsche Forschungsgemeinschaft  (Schm344/48-1), and  thanks  CONICET for continued support. 

\appendix
\section{Technical details}
The grand-canonical Monte Carlo method was pioneered by Torrie and Valleau \cite{tory}.
Our simulations were performed by a self-written program based on the book by Allen and Tildesley \cite{alden}.
Each Monte Carlo cycle consisted of one attempted insertion and one attempted destruction for each kind of particles, two attempted particle interchanges, and one  ordinary Monte Carlo attempted displacement
for each type.  At each potential, we started the simulations with a random
distribution of 50 particles of each type. 
One run consisted of between $10^6$ and $10^7$ cycles, depending on the fluctuations observed. Results for each case considered are based on an average of several runs --- since there is always a danger of the system getting stuck in a metastable state, we preferred to average over several runs rather than over one very long run. 

The simulations were performed for a tube of radius 5~a.u. (2.65~\AA) and a radius of 2.0~a.u. (1.06~\AA) for both ions. The radius refers to the radius of the effective position of the image charge. The results reported are for a tube of length 2000~a.u. (1058~\AA); thus, the maximum packing is 500~ions. Periodic boundary conditions were employed in the $z$-direction along the axis of the ring. We also performed simulations for tubes of 4000~a.u. length, and observed no differences in the behavior except larger fluctuations.

\ukrainianpart
\title{Накопичення заряду в нанотрубках: випадок 2-1 електроліту}
\author{В. Шміклер\refaddr{label1}, Д. Гендерсон\refaddr{label2}}
\addresses{
\addr{label1} Інститут теоретичної хімії, Університет Ульма, Німеччина
\addr{label2} Відділення хімії і біохімії, Університет Брігхама Янга, Прово, США
}

\makeukrtitle

\begin{abstract}
Ми розглядаємо 2-1 електроліт в контакті з вузькою нанотрубкою, яка дозволяє лише одновимірне накопичення вздовж осі.
Асиметрія не дозволяє  апріорі означити потенціал нульового заряду; натомість, природною точкою відліку є потенціал електрода,
біля  якого обидва іони мають 
той же електрохімічний потенціал;  значення останнього може служити для вимірювання іонофільності. Поблизу цього потенціалу іонофобні 
трубки наповнені розрідженим газом, іонофільні трубки є наповнені одномірним твердим тілом, що містить приблизно однакове число 
двовалентних іонів і моновалентних контріонів, стійкість структури забезпечується  сильним екрануванням кулонівської взаємодії 
індукованим зарядом на стінках трубки.
Наповнення трубки шляхом прикладання електродного потенціалу демонструє складну картину взаємодій між двома сортами іонів.

\keywords Монте Карло симуляції, електричний подвійний шар, іони, ємність, наноциліндрові пори
\end{abstract}

\end{document}